\documentstyle[12pt]{article}

\def\be{\begin{equation}}
\def\ee{\end{equation}}
\def\bea{\begin{eqnarray}}
\def\eea{\end{eqnarray}}
\def\beq{\begin{equation}}   \def\eeq{\end{equation}}
\def\bea{\begin{eqnarray}}   \def\eea{\end{eqnarray}}

\begin{document}
\begin{flushright}
UND-HEP-02-BIG\hspace*{.2em}05\\
TTP02-10 \\
hep-ph/0206261 \\
June 2002 \\
%\today \\
%cracow02v10.tex
%version 1.0
\end{flushright}

%\vspace{0.5cm}

\vspace{.3cm}
\begin{center} \Large 
{\bf 
{STUDYING CP VIOLATION  -- CHANCES FOR A NEW ALLIANCE}}
\footnote{Dedicated to Roberto Peccei on the occasion of his 60th 
birthday.}
\footnote{Invited talk given at {\em Meson 2002}, 
the 7th International Workshop on Meson Production, 
Properties and Interactions, 
Krakow, Poland, 
May 24 - 28, 2002.}
\\
\end{center}
\vspace*{.3cm}
\begin{center} {\Large 
I. I. Bigi }\\ 
\vspace{.4cm}
{\normalsize 
{\it Physics Dept.,
Univ. of Notre Dame du
Lac, Notre Dame, IN 46556, U.S.A.}
\footnote{Permanent address} \\
and \\
{\it Institut f\" ur Theoret. Teilchenp;hysik, 
Universit\" at Karlsruhe, D-???? Karlsruhe, 
FR Germany} }
\\
\vspace{.3cm}
{\it e-mail address: bigi.1@nd.edu } 
\vspace*{0.4cm}

{\Large{\bf Abstract}}
\end{center}
The last three years have seen a phase transition in our knowledge 
about CP violation: (i) Direct CP violation has been established in 
$K_L \to \pi \pi$. (ii) The first CP violation outside $K_L$ decays has
been  observed in $B_d \to \psi K_S$: 
$\langle {\rm sin}2\phi _1 \rangle = 0.78 \pm 0.08$ in amazing agreement 
with the prediction from the CKM description. The latter is thus seen 
as a tested theory. 
This increase in knowledge is not matched by progress in understanding. 
Searches for further manifestations of CP and T violation are mandatory. 
It is argued that one can profit greatly from studying atoms, molecules 
and nuclei.


%%%%%%%%%%%%
\tableofcontents 
%%%%%%%%%%
%%%%%%%%%%%%%% 
\section{Overview}
\label{INTRO}
%%%%%%%%%%%%

When I was invited to this meeting, I accepted for two 
reasons. The first one is Cracow: you only have to step onto the 
large square in the center of town to notice the beauty 
of the place and to realize you are in one of the truly great 
cities of Europe. The second reason is the Jagellonian University: 
over its long and illustrious history it has proven time and again that 
the power of the mind can overcome the force of the bayonets; therefore 
I always feel honoured by an association with this great university.

Once I started preparing this talk I realized there is a third 
reason why I am happy to talk to you, although most of you and I 
are usually running with different crowds: I think the time has 
come -- actually the opportunity has arisen -- to rethink the 
alignments that have developed in particle physics and come up 
with a new alliance to pursue the study of fundamental 
physics.

The message I want to communicate to you is the following: 
\begin{itemize}
\item 
In the last three years our knowledge about Nature's design 
has undergone such a large change that I would like to call 
it a phase transition: 
\begin{itemize}
\item 
We have witnessed the conclusion of an epoch with the  
experimental establishment of {\em direct} CP violation in the 
form of $\epsilon ^{\prime}/\epsilon \neq 0$. 
\item 
The first CP asymmetry outside the $K_L$ complex has been discovered  
in $B_d \to \psi K_S$. 
\item 
The CKM description of CP violation has been promoted from 
an {\em ansatz} to a {\em tested theory}. 
\item 
A clear whiff of {\em New Physics} has been 
caught in the evidence for neutrino oscillations. 
\end{itemize}
\item 
While our knowledge has increased, our 
understanding has not!
\item 
There  is added urgency to search for New Physics. One has  
a `King Kong' scenario for such a search and one where one has to 
rely on numerical precision, which represents a challenge 
not encountered before. 
\item 
A new alliance can provide new dimensions to such probes. 
\end{itemize}
%The remainder of the talk will be organized as follows: 
%after sketching the landscape of CP violation in 
%Sect.\ref{LAND}, I will emphasize the unreasonable success 
%of the CKM description in Sect.\ref{CKM}; after briefly 
%outlining the future of CP studies in Sect.\ref{FUTURE}, I will 
%make the case for a new alliance in Sect.\ref{NOVEL}.

%%%%%%%%%%%%%%%%%
\section{CP Violation -- The Landscape}
\label{LAND}
%%%%%%%%%%%%%%

%%%%%%%%%%%%%%
\subsection{Light flavours}
%%%%%%%%%%%%%

CP violation was discovered 1964 through the decay 
$K_L \to \pi ^+ \pi ^-$ -- causing considerable consternation 
among theorists \cite{LANDAU}. 
Till 1999, i.e. for 35 years, CP violation 
could be described by a {\em single} non-vanishing real number -- 
namely the phase between the quantities $M_{12}$ and 
$\Gamma _{12}$ in the $K^0 - \bar K^0$ mass matrix -- 
even in view of a large body of data! {\em Direct} CP violation 
has been unequivocally established in 1999. In the summer of 
2001 peaceful coexistence has been achieved between the data of 
NA48 and KTeV with a new world average \cite{BIINO}: 
\beq 
\langle \epsilon ^{\prime}/\epsilon _K \rangle = 
(1.72 \pm 0.18) \cdot 10^{-3} 
\eeq 
Quoting the result in this way does not do justice to the experimental 
achievement, since $\epsilon _K$ is a very small number itself. 
The sensitivity achieved and the control over systematic uncertainties 
established becomes more obvious when quoted in terms of actual 
widths:
\beq 
\frac{\Gamma (K^0 \to \pi ^+ \pi ^-) - 
\Gamma (\bar K^0 \to \pi ^+ \pi ^-)}
{\Gamma (K^0 \to \pi ^+ \pi ^-) + 
\Gamma (\bar K^0 \to \pi ^+ \pi ^-)} = 
(5.7 \pm 0.6) \cdot 10^{-6} \; !
\eeq
This represents a discovery of the very first rank -- 
{\em no} matter what theory does or does not say. The two 
groups deserve our deep respect, and they have certainly earned 
my admiration.

CPT symmetry implies that CP violation has to be accompanied by 
T violation; yet one would like to have a more direct manifestation 
of T violation. It has been demonstrated by CPLEAR through 
the Kabir test, where one compares
$K^0 \Rightarrow \bar K^0$ with 
$\bar K^0 \Rightarrow K^0$. To tag the initial state 
they rely on associated production of strangeness; the flavour 
identity of the final state is revealed through semileptonic 
decays. CPLEAR finds \cite{LEAR}
\beq 
A_T \equiv 
\frac{\Gamma(K^0 \Rightarrow \bar K^0) - 
\Gamma(\bar K^0 \Rightarrow K^0)}
{\Gamma(K^0 \Rightarrow \bar K^0) + 
\Gamma(\bar K^0 \Rightarrow K^0)} = (6.6 \pm 1.3 \pm 1.0) \cdot 
10^{-3} \neq 0
\eeq 
in full agreement with the value $(6.54\pm 0.24)\cdot 10^{-3}$ 
inferred from $K_L \to \pi\pi$.

The leading, namely linear term for the energy shift of a system inside a  
weak electric field $\vec E$ is described by a static 
quantity, the electric dipole moment $\vec d$ (EDM) 
\cite{BOOK}: 
\beq 
\Delta {\cal E} = \vec d \cdot \vec E + {\cal O}(E^2) 
\eeq
For a {\em non}-degenerate system with spin $\vec s$ one has 
$\vec d \propto \vec s$; therefore $\vec d \neq 0$ reveals 
T (and P) violation. The following upper bounds have been 
found for the EDM's of neutrons and electrons: 
\bea 
d_N &\leq& 6.3 \cdot 10 ^{-26} \; e cm \\
d_e &=& (-0.3 \pm 0.8) \cdot 10 ^{-26} \; e cm
\eea
These numbers reflect heroic experimental efforts that I cannot praise 
enough. The following comparisons might give you at least an intuitive 
feeling for the sensitivity achieved: the uncertainty in the 
electron's {\em magnetic} moment is about 
$2 \cdot 10^{-22}$ e cm and thus several orders of magnitude 
{\em larger} than the bound on its EDM! The bound on the neutron's 
EDM is smaller than its radius by 13 orders of magnitude. 
This corresponds to a relative displacement of an electron and a 
positron spread over the whole earth by less than 10 $\mu$ -- much 
less than the thickness of human hair!

%%%%%%%%%%%%%%%%%%%%%%
\subsection{CP violation in beauty decays}
%%%%%%%%%%%%%%%%%%

As predicted already 1980 the CKM description implies 
large CP asymmetries in several classes of $B$ decays involving 
$B_d - \bar B_d$ oscillations, most
notably  in $B_d \to \psi K_S$ \cite{CS,BS}. The existence of a huge CP
asymmetry  in $B_d \to \psi K_S$ has been established in 2001 by
BELLE and  BABAR \cite{BELLE1,BABAR1}; this spring they have presented
updates that agree very  nicely \cite{BELLE2,BABAR2}: 
\beq 
A_{CP}(B_d \to \psi K_S) =  
\left\{ 
\begin{array}{l}
0.75 \pm 0.09 \pm 0.04 \; \; {\rm BABAR} \\
0.82 \pm 0.12 \pm 0.05 \; \; {\rm BELLE}
\label{PHIDATA}
\end{array}  
\right.
\eeq
I have used here the notation $rate$ $(B_d(t) \to \psi K_S) 
\propto e^{-\Gamma _B t}
[1-A_{CP}(B_d \to \psi K_S){\rm sin}\Delta m_d ]$ 
\footnote{$t$ denotes the proper time of decay.} 
rather  than the CKM (near)identity 
$A_{CP}(B_d \to \psi K_S) = {\rm sin}2\phi _1$.  
The world average reads 
\beq 
\langle A_{CP}(B_d \to \psi K_S) \rangle = 
0.78 \pm 0.08 
\eeq
We can conclude 
that the CP asymmetry in $B_d \to \psi K_S$ is there for sure and it is 
huge -- as expected!

The data reveal directly that T and CP violation 
come together: 
\begin{itemize}
\item 
The function sin$\Delta m_dt$ representing CP 
violation is odd under $t \to - t$
and it describes  the data well. 
\item 
On the $\Upsilon (4S)$ neutral $B$ mesons are produced with their 
antiparticles, and one observes two $B$ decays with one, say a
semileptonic one, tagging the flavour identity and the other one  
being common to $B_d$ and $\bar B_d$ decays. I.e., one compares 
$e^+e^- \to (l^+X)_B +(\psi K_S)$ with 
$e^+e^- \to (l^-X)_{\bar B} +(\psi K_S)$ as a function of relative time
of  decay $\Delta t$; any difference constitutes a CP asymmetry. The data
show that the two distributions are shifted by the same amount, yet 
opposite direction away from $\Delta t =0$ with 
$\bar B^0 (t) \to \psi K_S$ proceeding sooner, i.e. faster. It should be
noted that EPR correlations 
\cite{EPR} are an essential element in this analysis: 
the two $B$ mesons are produced predominantly in a C-odd configuration; 
therefore even with oscillations they have to remain orthogonal to each 
other till one of them decays.

\end{itemize}

A new front has been opened up with studies of $B_d(t) \to \pi^+\pi^-$. 
The general parametrization for a CP asymmetry is given by 
\beq 
\frac{R_+(\Delta t) - R_-(\Delta t)}{R_+(\Delta t) + R_-(\Delta t)} = 
S {\rm sin}\Delta m_d \Delta t + C {\rm cos}\Delta m_d \Delta t 
\label{PIPIGEN}
\eeq
where $R_{[+,-]}(\Delta t)$ denotes the rate for 
$B_d\bar B_d \to ( l^+[l^-]X)_B + (\pi^+\pi^- )_B$ and 
\beq 
S = \frac{2{\rm Im}(q/p)\bar \rho (\pi\pi )}
{1+|(q/p)\bar \rho (\pi \pi )|^2} ,  
C= \frac{1-|(q/p)\bar \rho (\pi \pi )|^2}
{1+|(q/p)\bar \rho (\pi \pi )|^2} ,  
\bar \rho (\pi\pi) = 
\frac{T(\bar B_d \to \pi \pi)}{T( B_d \to \pi
\pi)}
\eeq
$C\neq 0$ constitutes {\em direct} CP violation -- yet so does 
$S \neq - A_{CP}(B_d \to \psi K_S)$!

This spring BELLE as well as BABAR have presented preliminary data 
on this channel \cite{BELLE2,BABAR2}: 
\bea 
BELLE: \; S_{\pi\pi} &=& - 1.21^{+0.38 + 0.16}_{-0.27-0.13} \; , \; 
C_{\pi\pi} =  0.94^{+0.25}_{-0.31} \pm 0.09 \\
BABAR: \; S_{\pi\pi} &=& - 0.01 \pm 0.37 \pm 0.07 \; , \; 
C_{\pi\pi} =  0.02 \pm 0.29 \pm 0.07
\eea
Taken at face value the two sets would suggest very different
messages, namely BELLE seeing a large CP asymmetry also of the {\em
direct}  variety, while BABAR not seeing any CP asymmetry. In view of the
large  statistical uncertainties, such conclusions would be premature. 
Yet there is another somewhat more subtle point I would like to 
emphasize 
\cite{ELEPHANT}. The first order should be to analyse whether one has
observed  something that goes beyond what has been established in 
$B_d (t) \to \psi K_S$. There one has found a large complex phase 
that will contribute to $B_d (t) \to \pi ^+ \pi ^-$ as well in the 
form of 
\beq 
S_{\pi\pi} = - S_{\psi K_S} = - (0.75 \div 0.82) \; , \; 
C_{\pi\pi} = 0
\label{MIN}
\eeq   
The relevant question is {\em not} whether Eq.(\ref{MIN}) yields the 
best fit to the data -- far from it! What matters is whether it 
provides an {\em acceptable} description. If not, one has established
{\em direct} CP violation  in $B_d$ decays, since two  
flavour non-specific decay channels have exhibited different CP 
asymmetries. Not observing any CP asymmetry in 
$B_d (t) \to \pi ^+ \pi ^-$ would actually establishes large direct CP 
violation!

BELLE has shown some intriguing evidence for a large CP asymmetry 
in $B^{\pm} \to \pi ^{\pm} K_S$: 
\beq 
A_{CP} = 0.46 \pm 0.15 \pm 0.02
\eeq
CPT invariance requires such an effect to be compensated by an 
asymmetry in channels that can rescatter with $\pi^{\pm}K_S$; 
$K^{\pm}\pi ^0/\eta$ should figure prominently among such channels. 
Finding indeed the compensating effect there would be a very 
persuasive cross check.

%%%%%%%%%%%%%%%%%
\section{The Unreasonable Success of the CKM Description}
\label{CKM}
%%%%%%%%%%%%%%%
 
Due to CPT symmetry CP violation can be implemented only 
through complex phases; those can arise in charged current 
couplings. CP invariance can thus be broken only if the dynamical 
substrate is sufficiently complex. Kobayashi and Maskawa realized 
in 1973 that within the SM this requires three families. Denoting 
the unitary matrices diagonalizing the "up" and "down" quark mass 
matrices by $T^U$ and $T^D$, respectively, one has for the 
CKM matrix $V_{CKM} = T_L^U (T_L^D)^{\dagger}$; i.e., 
$V_{CKM}$ represents a {\em non}trivial unitary matrix, 
{\em unless} the 
"up" and "down" quark sectors are {\em aligned}. CKM parameters are 
thus intrinsically connected with the mass generation for quarks. 
The unitarity conditions for the $3\times 3$ matrix $V_{CKM}$ yield  
three weak universality relations and six triangles in the 
complex plane. Among the latter is one that involves $b$ quark 
couplings and has naturally large 
angles.

The status of the KM pre- \& post-dictions in '98 was as
follows: 
\begin{itemize}
\item 
$\Delta m_K$, $\epsilon _K$ and $\Delta m(B_d)$ can be reproduced within
a factor of two. 
\item 
Cancellations reduce $\epsilon ^{\prime}/\epsilon _K$ below its 
`natural' value of $10^{-3}$; yet there were some heretics objecting 
to this certitude \cite{TRIESTE}. 
\item 
Concerning the CP asymmetries in $B$ decays it was stated that 
some have to be of order unity with no "plausible" deniability; 
in the early '90's -- i.e. before the discovery of top 
quarks -- this was specified to predicting 
sin$2\phi_1[\beta] \sim 0.6 - 0.7$ if today's estimates of 
$f_B$ are used \cite{BEFORETOP}. 
\item 
In '98 courageous souls predicted  
sin$2\phi_1[\beta] \sim 0.72 \pm 0.07$ \cite{ACHILLE}.

\end{itemize}
It is indeed true that large fractions of $\Delta m_K$, 
$\epsilon _K$ and $\Delta m_B$ and even most of 
$\epsilon ^{\prime}$ could be due to New Physics; constraints 
from data thus translate into `broad' bands in plots of the 
unitarity triangle. Yet such a statement seemingly reflecting 
facts misses the deeper message! Consider a map of Krakow and plot 
where participants of this conference and their companions can be 
found during a day. You will find rather broad bands 
covering at least the area between the Wawel, Kazimierz, the 
Barbikane and Bursa Pigonia on Garbarska street. Yet the real 
point is that so many physicists from all 
over the world are concentrated in Krakow. This cannot be an 
accident; there has to be a good reason --  and there is, of course! 
Similarly one has to keep in mind that 
the dimensional quantities describing the weak observables 
span several orders of magnitude. It is highly remarkable that 
the CKM description can always get to within a factor of two 
or three -- in particular with numerical values in its parameters 
and the fermion masses that {\em a priori} would have seemed 
to represent frivolous choices, like $m_t \simeq 180$ GeV. 
And it appears right on the mark for sin$2\phi_1$! Hence I 
conclude that the CKM description is no longer a mere ansatz, but 
a tested -- though not confirmed -- theory; its forces are with us to
stay.  En passant we have learnt that when complex phases surface they 
can be large.

An aside might be allowed here. 
A CP odd quantity depends also on the
$sin$  (and $cos$) of the three CKM angles in addition to the complex 
phase. While the latter is large, the former are small or even 
tiny -- something that can be understood in the context of theories 
with extra dimensions \cite{MARCH} -- allowing CP violation a 
la CKM to be generated perturbatively \cite{DENT}. A large 
CP {\em asymmetry} can arise when also the decay rate is 
suppressed by small CKM parameters as it happens in $B$ decays.

Yet this new and spectacular success of the SM does not resolve 
any of its mysteries -- why are there families, why three, 
what is the origin of the peculiar pattern in the quark mass matrices 
-- they actually deepen them.  
Consider the structure of the CKM matrix: 
\beq 
|V_{CKM}| \sim 
\left( 
\begin{array}{ccc} 
1 & {\cal O}(\lambda ) & {\cal O}(\lambda ^3) \\ 
{\cal O}(\lambda ) & 1 & {\cal O}(\lambda ^2) \\ 
{\cal O}(\lambda ^3) & {\cal O}(\lambda ^2) & 1 
\end {array} 
\right) 
, \; \lambda = {\rm sin}\theta _C
\label{DRES}
\eeq
There has to be fundamental information encoded in this 
hierarchical pattern. 
The situation can be characterized by saying "we know so much, yet 
understand so little!" It emphasizes that the SM is 
incomplete, that New Physics must exist.

%%%%%%%%%%%%%%%%%%%%%
\section{On the Future of CP Studies}
\label{FUTURE}
%%%%%%%%%%%%%

%%%%%%%%%%%%%
\subsection{The `King Kong' scenario}
%%%%%%%%%%%%%%

$\Delta S=1,2$ dynamics have provided several examples of revealing the 
intervention of features that represented New Physics 
{\em at that time} like parity and CP violation, the 
existence of quark families etc.  
It thus has been instrumental in the evolution of the SM. This happened 
through the observation of `qualitative' discrepancies; i.e., 
rates that were expected to vanish did not, or rates were 
found to be smaller than expected by 
{\em several orders of magnitude}. 
Such an indirect search for New Physics can be characterised as a 
`King Kong' scenario: "One might be unlikely to encounter King 
Kong; yet once it happens there can be no doubt that one has 
come across something extra-ordinary". Such a situation 
can be realized again in different ways: 
\begin{itemize}
\item 
Dedicated searches for EDM's of neutrons, atoms and molecules are 
a definite must \cite{BOOK,POSP} -- no excuses are 
acceptable. For one should keep in mind that they are so tiny in the 
SM for reasons very specific to the CKM implementation of T violation.  
\item 
Searching for a transverse polarization of muons in 
$K^+ \to \mu ^+ \nu \pi ^0$ is a promising way to uncover the 
intervention of Higgs based CP violation. 
\item 
There exists a large literature on $D^0 - \bar D^0$ oscillations 
with predictions covering several orders of 
magnitude. That does not mean that they are all equally credible, 
though. A systematic analysis has been given now based on the 
operator product expansion expressing $x = \Delta m_D/\Gamma$ and 
$y=\Delta \Gamma/2\Gamma$ in powers of $1/m_c$, $m_s$ and KM factors. 
One finds \cite{DOSC} $x$, $y$ $\sim {\cal O}(10^{-3})$ with the 
prospects for reducing the uncertainties rather slim; one should also 
note that $y$ is more sensitive to violations of quark-hadron duality 
than $x$. Recent claims \cite {LIGETI} that there is a model independant 
estimate yielding $x$, $y$ around 1\% are greatly overstated.

After the early indication from FOCUS that $y$ might be around 1\%, 
other data have not confirmed it 
\cite{PUROHIT}: $y$ as well as $x$ are consistent with zero on the 
1-2 \% level. Detailed searches for CP asymmetries in $D$ decays 
have been and are being undertaken: data are consistent with no 
asymmetries so far, again on the few percent level \cite{STENSON}.

While it is possible to construct New Physics scenarios producing 
effects as large as 10 \% or so in charm transitions, a more reasonable
range is the  1\% level. Thus one is {\em only now} entering territory
where there are some  realistic prospects for New Physics to emerge. 
\end{itemize}

%%%%%%%%%%%%%%%
\subsection{The `Novel Challenge'}
%%%%%%%%%%%%%

The situation is quite different in $B$ transitions since the 
CKM dynamics already generate large CP asymmetries. The one 
significant exception arises in 
$B_s(t) \to \psi + \eta/\phi$ where one can reliably predict a small
asymmetry  not exceeding 2 \% for reasons that are very specific to the 
CKM description \cite{BS}; anything beyond that is a manifestation 
of New Physics.

We can expect that the $B$ factories, 
BTeV and LHC-b will allow to measure a host of CP asymmetries 
with experimental uncertainties not exceeding a few percent. 
The question then arises whether we can exploit this 
level of sensitivity {\em theoretically}; i.e., whether one can 
interprete data and make predictions with no more than a 
few percent theoretical uncertainty. I do not consider such a goal for
theoretical  control a luxury. With the exception of 
$B_s(t) \to \psi +\phi/\eta$ noted above CP asymmetries in $B$ decays
often  are large within the SM (or are severely restricted by the 
need for strong phase shifts). Therefore New Physics typically cannot
change 
SM predictions by orders of magnitude. Furthermore I had argued 
above that
the success  of the CKM theory in describing weak observables
characterized by  scales ranging over several orders of magnitude is
highly non-trivial.  Accordingly I do not find it very likely that New
Physics, i.e. dynamics {\em not} noted before, will  affect transition
rates for
$B$ hadrons in a {\em massive} way. 
Thus one is faced with a novel challenge: can one be 
confident of having established the presence of New Physics when 
the difference between the expected and the observed signal is much 
less than an order of magnitude? To be more specific: assume 
one predicts an asymmetry of 40 \%, yet observes 60 \% -- can one 
be certain of New Physics? What about if one observes 50\%? 
Interpreting such {\em quantitative} 
discrepancies represents a challenge 
which we have not faced before.

%%%%%%%%%%%%
\section{The New Alliance}
\label{NOVEL}
%%%%%%%%

Limitations of T and CP invariance are probed in a large number of 
dynamical environments 
using a host of experimental techniques and set-ups: one 
searches 
for  EDM's of neutrons, electrons, atoms and molecules; for CP 
asymmetries  
in the decays of muons, 
$\tau$ leptons and hadrons with strangeness, charm and beauty quantum 
numbers; one performs both fixed target and
colliding  beam experiments; one harnesses the coldest temperatures 
ever achieved on earth -- when producing ultracold neutrons -- 
and the highest temperatures witnessed in the universe since just after 
the big bang -- when running at the LHC. Every high energy 
physics lab has a dedicated program of CP studies; several 
labs of nuclear and atomic physics participate in a 
significant way. Clearly whenever so
many so  diverse groups of researchers pursue a goal, that goal must be
of great importance.

In particle physics fundamental dynamics is probed on three different 
frontiers, namely the high energy, high sensitivity and high precision 
frontier. This is a very useful and natural classification and division 
of labour. However we should not allow it to become a straightjacket. 
It is about time -- maybe even high time -- to form new alliances 
to probe fundamental dynamics. T and CP studies are a prime example 
with a well defined -- though not widely perceived -- program 
for such a new alliance: 
\begin{itemize}
\item 
Such studies provide an extremely sensitive low energy probe of
fundamental  dynamics. 
\item 
We know that the CKM mechanism cannot generate the baryon number of the 
Universe. 
\item 
The {\em non-}CKM dynamics thus required could easily be buried in 
beauty decays under the `background' of large CKM CP asymmetries in 
an example of Telegdi's famous dictum that `yesterday's sensation 
is today's calibration and tomorrow's background'. 
\item 
At the same time such dynamics would have hardly a competition from CKM 
forces when probing T and CP invariance in light flavour systems. 
\item 
Furthermore we would benefit greatly from the expertise developed in 
other areas of physics and the opportunities offered by such 
complex systems like nuclei and molecules as labs to search
for  T {\em odd} effects. One should keep in mind that the emphasis 
here is on sensitivity, rather than precise theoretical control, as 
long as there is no competition from known physics. 
\item 
Many such experiments are truly `table top' set-ups, though of a 
highly sophisticated kind \cite{LAM}. 
\end{itemize}

%%%%%%%%%%%%%
\section{Summary}
%%%%%%%%%%%

"The SM is consistent with the data" is a statement most of you 
experience as a worn-out refrain. However in the last two years it has 
acquired new dimensions (pun intended) 
leaving the Higgs sector as the only remaining  
`terra incognita' of the SM. For an essential test of the 
CKM description of CP violation has been performed in $B \to \psi K_S$; 
the first CP asymmetry outside the $K^0 - \bar K^0$ complex has 
been observed, and it is huge -- as expected! 
In my judgement the CKM description of 
CP violation thus has been promoted from an ansatz to a tested theory 
that is going to stay with us. Yet this success of the CKM theory 
does not resolve any of the central mysteries of the SM concerning the 
heavy flavour sector: why is there family replication, why are there 
three families, what generates the very peculiar pattern in the 
quark masses and the CKM parameters? It actually 
deepens those mysteries and -- in my view -- makes a convincing 
case that the SM is incomplete!

This conclusion is further strengthened by three observations: 
\begin{enumerate} 
\item 
Strong experimental evidence has been accumulated for 
neutrino oscilations. 
One might argue that those can be 
incorporated into a `trivial' extension of the SM by just 
adding right-handed neutrinos without gauge interactions; one 
can engineer neutrino Yukawa couplings to Higgs doublets 
in such a way as
to  obtain the needed mass matrices. However that would be highly 
contrived; the only known {\em natural} way to understand the 
tiny neutrino masses is  
through the see-saw mechanism,
which  requires Majorana masses. Yet those cannot be obtained from
doublet  Higgs fields. While the see-saw mechanism suggests a highly  
hierarchical structure in the neutrino parameters, this is not
necessarily so as pointed out by Jezabek, since there are 
actually two matrices describing $\nu$ mass-related parameters 
\cite{JEZABEK}.  
\item 
The `strong CP problem' remains unsolved \cite{PECCEI,BOOK}. 
\item 
We know now that CKM dynamics cannot generate 
the baryon number of the Universe.    
\end{enumerate} 
It is obvious then that the dedicated study of heavy flavour dynamics 
can never become marginal, let alone obsolete. 
Much more can be said about this; here I want to comment only on 
directions for CP studies. It hardly needs justification to analyze 
all kinds of CP asymmetries in the decays of beauty hadrons with as 
much precision as possible. Yet this truth should not make us forget 
about other important avenues to pursue. For the 
{\em non}-CKM CP violating dynamics needed to generate the Universe's 
baryon number have a much better chance to reveal themselves through 
their impact on T {\em odd} effects for light flavour hadrons 
and leptons. At the same time we would benefit 
tremendously from the expertise accumulated and the opportunities 
spotted in different areas: atoms, molecules and nuclei can 
represent promising labs to search for $T$ {\em odd} effects like 
EDM's etc.

\section*{Acknowledgments}

I would like to thank the organizers for creating such an  
enjoyable meeting.  
This work has been supported by the NSF under the grant  
PHY-0087419.

\end{document}